%% file: root.tex
\documentclass[letterpaper, 10 pt, conference]{ieeeconf}  

\IEEEoverridecommandlockouts                              
\usepackage{mathtools,amssymb,commath}
\usepackage{amsmath}
\usepackage{graphicx,import}
\usepackage{hyperref}
\hypersetup{colorlinks=true}   

\DeclareMathOperator{\trace}{tr} 
\DeclareMathOperator{\vect}{vec}
\DeclareMathOperator{\rank}{rk} 
\DeclareMathOperator{\real}{Re}

\usepackage{mathtools,amsthm}

\title{\LARGE \bf
Automatic Gain Tuning  of a Momentum Based \\Balancing Controller for Humanoid Robots\\
}

\author{Daniele Pucci, Gabriele Nava and Francesco Nori$^{1}$
\thanks{*This paper was supported by the FP7 EU project KoroiBot (No. 611909 ICT 2013.10 Cognitive Systems and Robotics)}
\thanks{$^{1}$ The authors are with the iCub Facility department, Istituto Italiano di Tecnologia,
        Via Morego 30, Genoa, Italy
        {\tt\small name.surname@iit.it}}%
}

\input{definitions}

\begin{document}

\maketitle
\thispagestyle{empty}
\pagestyle{empty}

\begin{abstract}
    
    This paper proposes a technique for automatic gain tuning of a momentum based balancing controller for humanoid robots.
    The controller ensures the stabilization of the centroidal dynamics and the associated zero dynamics.
    Then, the closed-loop, constrained joint space dynamics is linearized and the controller's gains are chosen so as to obtain
    desired  properties of the linearized system. 
    Symmetry and positive definiteness constraints of gain matrices are enforced by proposing a tracker for symmetric positive definite matrices.
    Simulation results are carried out on the humanoid robot iCub. 
\end{abstract}

\input{tex/introduction}
\input{tex/background}
\input{tex/balancing}

\input{tex/tuning}
\input{tex/results}

\input{tex/conclusions}

\addtolength{\textheight}{0cm}     

\import{tex/}{appendix}
\bibliographystyle{IEEEtran}
\bibliography{IEEEabrv,BibliographyHUM}

\end{document}

%% file: definitions.tex
\newtheorem{lemma}{\bf{Lemma}}

\DeclareMathOperator*{\argmin}{argmin}

\makeatletter
\newsavebox\myboxA
\newsavebox\myboxB
\newlength\mylenA
\newcommand*\xoverline[2][0.75]{%
    \sbox{\myboxA}{$\m@th#2$}%
    \setbox\myboxB\null
    \ht\myboxB=\ht\myboxA%
    \dp\myboxB=\dp\myboxA%
    \wd\myboxB=#1\wd\myboxA
    \sbox\myboxB{$\m@th\overline{\copy\myboxB}$}
    \setlength\mylenA{\the\wd\myboxA}
    \addtolength\mylenA{-\the\wd\myboxB}%
    \ifdim\wd\myboxB<\wd\myboxA%
       \rlap{\hskip 0.5\mylenA\usebox\myboxB}{\usebox\myboxA}%
    \else
        \hskip -0.5\mylenA\rlap{\usebox\myboxA}{\hskip 0.5\mylenA\usebox\myboxB}%
    \fi}
\makeatother

%% file: tex/introduction.tex
\section{Introduction}
\label{sec:introduction}

Humanoid robotics is an undoubtedly   flourishing field of research. Locomotion and manipulation have received a special attention from the control community, and the results shown at the DARPA robotics challenge are both stimulating and promising~\cite{koolen2015design,Hopkins2015a}.  
Despite these advances, robust controllers for balancing and walking  of humanoids still require a special focus of the robotics community. Furthermore, when these controllers are implemented on real platforms, the achievement of desired system performances usually requires time consuming tuning of the (often very numerous) gains characterizing the control laws. 
This paper proposes a technique to tune automatically the gains of a momentum based balancing strategy for humanoid robots.

A classical approach to the modelling and control of humanoid robots is based on considering the robot attached to ground, i.e. the robot is considered to be \emph{fixed-base}~\cite{Quiang2000}. In this case, in fact, well-known classical control strategies for manipulators can be directly applied for robot control. The limitations of this approach arise when attempting to tackle the general control problem of a humanoid interacting with its surrounding environment. For instance, running involves flight phases where the fixed-base approach clearly fails.  
At the modelling level, the Euler-Poincar\`{e} equations provide singularity free equations of motion for the humanoid robot~\cite[Chapter 13]{marsden2013introduction}, and efficient algorithms can be applied for determining the components of these equations \cite{Featherstone2007}. 
When considering these equations of motion, the mechanical system representing the humanoid robot is usually under actuated, and this forbids the full feedback linearization of the closed-loop system 
\cite{Acosta05,Featherstone2007}. The system underactuation is usually dealt with by means of
 constraints that arise from the contacts between the robot and the environment.
This requires a close attention to the forces the robot exerts on the environment since uncontrolled forces may break the contacts, thus rendering the control of the robot critical. To ensure that the contact forces respect some physical 
constraints, different optimization procedures can be applied \cite{Wensing2013,Kuindersma2014}.

\emph{Task-based} control strategies have proven to be an efficient solution for balancing and walking of humanoid robots \cite{Stephens2010,Herzog2014,Frontiers2015}.
The aim of these strategies is the achievement of several control objectives, which are organized in a hierarchical structure. The possibility of defining different
control objectives with different
priorities is an efficient way to deal with manipulation tasks while balancing \cite{farnioli2015}. Furthermore, high priority tasks may be used to  control directly the 
contact forces the robot exerts at the contact points, e.g. through the control of the center of mass dynamics \cite{Lee2012}. 

When the above control algorithms are implemented in real applications, a long and tedious tuning of control gains is often required to achieve desired system performances. 
Despite the large number of gain optimization procedures for classical dynamical systems (see, e.g., \cite{Tesh2002,Narong2009}) gain optimization techniques for floating base systems, and in particular in the field 
of humanoid robots, still needs more investigations.
Preliminary results in this direction consist in 
applying classical LQR approaches to the linearized humanoid robot dynamics \cite{manson2014,Alonso2016}. In particular, LQR based optimization techniques can be applied to the so-called centroidal dynamics~\cite{manson2014}. Another approach may consist of considering simplified robot models, and then apply classical gain scheduling procedures for balancing purposes~\cite{Xing2010}.

In this paper, we propose a gain tuning method for the momentum-based control strategy we implemented on the iCub humanoid robot~\cite{nava2016}. Contrary to~\cite{manson2014}, we optimize the gains by imposing desired properties of the linearized joint space dynamics. The approach handles any number of contacts between the robot and the environment.
Symmetry and positive definiteness constraints of gain matrices are enforced via a nonlinear tracker for symmetric positive definite matrices. Simulation results verify the presented approach.

The paper is organized as follows. Section~\ref{sec:background} introduces notation and  system modelling. Section~\ref{sec:balancing} recalls and complements the momentum-based control strategy~\cite{nava2016}. Section~\ref{sec:tuning} presents the linearization and the gain optimization
procedure. Section~\ref{sec:results} presents simulations results using the iCub robot model. Conclusions and future works conclude the paper. 

%% file: tex/background.tex
\section{BACKGROUND}
\label{sec:background}

\subsection{Notation}
\begin{itemize}
\item $\mathcal{I}$ defines an inertial frame of reference, with its $z$ axis pointing against the gravity. The constant $g$ denotes the norm of the gravitational acceleration.
\item Given a matrix $A \in \mathbb{R}^{m \times n}$, we denote with $A^{\dagger}\in \mathbb{R}^{n \times m}$ its Moore Penrose pseudoinverse. 
\item $e_i \in  \mathbb{R}^m$ is the canonical vector, consisting of all zeros but the $i$-th component that is equal to one.
\item We denote with $m$ the total mass of the robot.
\end{itemize}

\subsection{Modelling}
The robot is modelled as a multi-body system composed of $n + 1$ rigid bodies, called links, connected by $n$ joints with one degree of freedom each. We also assume that
none of the links has an \emph{a priori} constant pose with respect to an inertial frame, i.e. the system is \emph{free floating}.

The robot configuration space is the Lie group $\mathbb{Q} = \mathbb{R}^3 \times SO{(3)} \times \mathbb{R}^n$  and it is characterized by the \emph{pose} 
(position and orientation) of a base frame attached to a robot's link, and the joint positions.
An element $q \in \mathbb{Q}$ can be defined as the following triplet: $q = (\prescript{\mathcal{I}}{}p_{\mathcal{B}}, \prescript{\mathcal{I}}{}R_{\mathcal{B}}, q_j)$ 
where $\prescript{\mathcal{I}}{}p_{\mathcal{B}} \in \mathbb{R}^3$ denotes the position of the base frame with respect to the inertial frame,
$\prescript{\mathcal{I}}{}R_{\mathcal{B}} \in \mathbb{R}^{3\times3}$ is a rotation matrix representing the orientation of a \emph{base frame}, and $q_j \in \mathbb{R}^n$ is the 
joint configuration. 
The velocity of the multi-body system can be characterized by the \emph{algebra} of the group 
defined as $\mathbb{V} = \mathbb{R}^3 \times \mathbb{R}^3 \times \mathbb{R}^n$.
An element of $\mathbb{V}$ is a triplet $\nu = ( ^\mathcal{I}\dot{ p}_{\mathcal{B}},^\mathcal{I}\omega_{\mathcal{B}},\dot{q}_j) = (\text{v}_{\mathcal{B}}, \dot{q}_j)$,
where $^\mathcal{I}\omega_{\mathcal{B}}$ is the angular velocity of the base frame expressed w.r.t. the inertial frame, 
i.e. $^\mathcal{I}\dot{R}_{\mathcal{B}} = S(^\mathcal{I}\omega_{\mathcal{B}})^\mathcal{I}{R}_{\mathcal{B}}$.
A more detailed description of the floating base model is provided in \cite{nava2016}.

We assume that the robot interacts with the environment by exchanging $n_c$ distinct wrenches. The equations of motion of the multi-body system can be described
applying the Euler-Poincar\'e formalism \cite[Ch. 13.5]{Marsden2010}:
\begin{align}
\label{eq:system}
{M}(q)\dot{{\nu}} + {C}(q, {\nu}){\nu} + {G}(q) =  B \tau + \sum_{k = 1}^{n_c} {J}^\top_{\mathcal{C}_k} f_k
\end{align}
where ${M} \in \mathbb{R}^{n+6 \times n+6}$ is the mass matrix, ${C} \in \mathbb{R}^{n+6 \times n+6}$ is the Coriolis matrix, ${G} \in \mathbb{R}^{n+6}$ is the gravity 
term, $B = (0_{n\times 6} , 1_n)^\top$ is a selector matrix, $\tau \in \mathbb{R}^{n}$ is a vector representing the internal actuation torques, and 
$f_k \in \mathbb{R}^{6}$ denotes an external wrench applied by the environment to the link of the $k$-th contact. The Jacobian 
${J}_{\mathcal{C}_k} = {J}_{\mathcal{C}_k}(q)$ is the map between the robot's velocity ${\nu}$ and the linear and angular velocity at the $k$-th contact link.

As described in \cite[Sec. 5]{traversaro2016}, it is possible to apply a coordinate transformation in the state space $(q,{\nu})$ that transforms the system dynamics~\eqref{eq:system} into a new form where
the mass matrix is block diagonal, thus decoupling joint and base frame accelerations. Also, in this new set of coordinates,  the first six rows of Eq. \eqref{eq:system} are the \emph{centroidal dynamics}\footnote{In the specialized literature, the term \emph{centroidal dynamics} 
is used to indicate the rate of change of the robot's momentum expressed at the center-of-mass, which then equals the summation of all external wrenches acting on the 
multi-body system \cite{orin2013}.}. As an abuse of notation, we assume that system \eqref{eq:system} has been transformed in this new set of coordinates, i.e. 
\begin{IEEEeqnarray}{RCL}
\label{centrTrans}
M(q) &=& \begin{bmatrix} {M}_b(q) & 0_{6\times n} \\ 0_{n\times 6} & {M}_j(q) \end{bmatrix}, \quad
H    = M_b \text{v}_{\mathcal{B}},
\end{IEEEeqnarray}
where ${M}_b \in \mathbb{R}^{6\times 6}, {M}_j \in \mathbb{R}^{n\times n}$, ${H}:=(H_L,H_\omega)\in \mathbb{R}^6$ is the robot momentum, and $H_L, H_\omega \in \mathbb{R}^3$ are the linear and angular momentum at the center of mass, respectively.
                           
Lastly, it is assumed that a set of holonomic constraints acts on System \eqref{eq:system}. These holonomic constraints are of the form $c(q) = 0$, and may represent,
for instance, a frame having a constant pose w.r.t. the inertial frame.
In the case where this frame corresponds to the location at which a rigid contact occurs on a link, we represent the holonomic constraint as
$
{J}_{\mathcal{C}_k}(q) {\nu} = 0.$
Hence, the kinematic constraint associated with all the rigid contacts can be represented as
\begin{IEEEeqnarray}{RCL}
\label{eqn:constraintsAll}
{J}(q) {\nu} {=} 
\begin{bmatrix}{J}_{\mathcal{C}_1}(q) \\ \cdots \\ {J}_{\mathcal{C}_{n_c}}(q)  \end{bmatrix}{\nu}  {=} 
\begin{bmatrix} J_b & J_j  \end{bmatrix}{\nu}  
&=& J_b {\text{v}}_{\mathcal{B}}+ J_j \dot{q}_{j} = 0,
\IEEEeqnarraynumspace
\end{IEEEeqnarray}
with $J_b \in \mathbb{R}^{6n_c \times 6},J_j \in \mathbb{R}^{6n_c \times n} $, and  $\dot{q}_j \in \mathbb{R}^n$ the joint space velocity.
The base frame velocity is denoted by $\text{v}_{\mathcal{B}} \in \mathbb{R}^6$, which in the new coordinates yielding a block-diagonal mass matrix is given by $\text{v}_{\mathcal{B}} = (\dot{p}_c,\omega_o)$, where $\dot{p}_c \in \mathbb{R}^3$ is the velocity of the system's center of mass ${p}_c \in \mathbb{R}^3$, and $\omega_o \in \mathbb{R}^3$ is the so-called system's \emph{average angular velocity}.
By differentiating the kinematic constraint Eq. \eqref{eqn:constraintsAll}, one obtains
\begin{equation}
    \label{eq:constraints_acc}
   J\dot{\nu}+\dot{J}\nu = J_b \dot{\text{v}}_{\mathcal{B}}+ J_j \ddot{q}_{j} + \dot{J}_b {\text{v}}_{\mathcal{B}}+ \dot{J}_j \dot{q}_{j} = 0.
\end{equation}

%% file: tex/balancing.tex
\section{RECALLS AND COMPLEMENTS ON THE MOMENTUM-BASED CONTROL STRATEGY}
\label{sec:balancing}

We recall and complement  the \emph{momentum-based} control strategy implemented on our iCub humanoid robot~\cite{nava2016}. The control objective is the stabilization of a desired robot momentum and the stability of the associated zero dynamics. 

\subsection{Momentum Control}
Recall that the rate-of-change of the robot momentum equals the net external wrench acting on the robot, which in the present case reduces to the contact
wrench $f:=(f_1,\cdots,f_{n_c}) \in \mathbb{R}^{6n_c} $ plus the gravity wrenches. Then, in view of Eq. \eqref{centrTrans}, the rate-of-change of the robot momentum can be expressed as:
\begin{IEEEeqnarray}{RCCCL}
\label{hDot}
 \frac{\dif }{\dif t}(M_b {\text{v}_\mathcal{B}}) &=& \dot{H}(f) &=& J_b^{\top}f - mge_3,
\end{IEEEeqnarray}
where $e_3 \in \mathbb{R}^6$.

The control objective is defined as the stabilization of a desired robot momentum $H^d \in \mathbb{R}^6 $. Let $\tilde{H} = H - H^d$ define the momentum error. Assuming  that the contact wrenches $f$ can be chosen at will, then we choose $f$ such that \cite{nava2016}:
\begin{IEEEeqnarray}{RCL}
    \label{hDotDes}
    \dot{H}(f) &=& 
    \dot{H}^* := \dot{H}^d - K_p \tilde{H} - K_i I_{\tilde{H}}    
    \IEEEeqnarraynumspace  \IEEEyessubnumber \label{eq:dotH_f}  \\
    \dot{I}_{\tilde{H}} &=&
                     \begin{bmatrix} 
                     {J}_{G}^L(q_j) \\ 
                     {J}_{G}^{\omega}(q^d_j)
                     \end{bmatrix}\dot{q}_j \IEEEyessubnumber  \label{eq:IhTilde} 
\end{IEEEeqnarray}
$K_p, K_i {\in} \mathbb{R}^{6\times 6}$ two symmetric, positive definite matrices and 
\begin{IEEEeqnarray}{RCL}
\label{eqn:jacobian} 
\bar{J}_G(q_j) &{:=}& - M_bJ^{\dagger}_bJ_j =  \begin{bmatrix} {J}_{G}^L(q_j) \\ {J}_{G}^{\omega}(q_j)\end{bmatrix} \in \mathbb{R}^{6\times n}, {J}_{G}^L, {J}_{G}^{\omega}\in \mathbb{R}^{3\times n}\IEEEeqnarraynumspace \nonumber
\end{IEEEeqnarray} 
If $n_c > 1$, there are infinite contact wrenches $f$ that satisfy Eq.~\eqref{eq:dotH_f}.
We parametrize the set of solutions $f$ to \eqref{eq:dotH_f} as:
\begin{equation}
    \label{eq:forces}
    f = f_1 + N_{b}f_0
\end{equation}
with $f_1 =  J_b^{\top\dagger} \left(\dot{H}^*+ mg e_3\right)$, $N_b \in \mathbb{R}^{6n_c \times 6n_c}$ the projector into the null space of $J_b^{\top}$, and $f_0\in \mathbb{R}^{6n_c}$ the wrench redundancy that does not influence $\dot{H}(f) =
    \dot{H}^*$.
To determine the control torques that instantaneously realize the contact wrenches given by \eqref{eq:forces}, we use the dynamic 
equations \eqref{eq:system} along with the constraints \eqref{eq:constraints_acc}, which yields:
\begin{equation}
    \label{eq:torques}
    \tau = \Lambda^\dagger (JM^{-1}(h - J^\top f) - \dot{J}\nu) + N_\Lambda \tau_0
\end{equation}
with $\Lambda = {J_j}{M_j}^{-1} \in \mathbb{R}^{6n_c\times n}$,  $N_\Lambda \in \mathbb{R}^{n\times n}$  the projector onto the nullspace
of $\Lambda$, the vector $h:={C}(q, {\nu}){\nu} + {G}(q)  \in \mathbb{R}^{n+6}$,  and $\tau_0 \in \mathbb{R}^n$  a free variable.

\subsection{Stability of the Zero Dynamics}
The stability of the zero dynamics is attempted by means of a so called \emph{postural task}, which exploits the free variable $\tau_0$.
Partition the vector $h$ as follows: $h = (h_b,h_j), h_b\in\mathbb{R}^6,  h_j\in\mathbb{R}^n$. A choice of the postural 
task that ensures the stability of the zero dynamics on one foot is \cite{nava2016}:
\begin{IEEEeqnarray}{lCr}
\label{posturalNew}
    \tau_0 &=& h_j - J_j^\top f + u_0
\end{IEEEeqnarray}
where 
$u_0 := -K^j_{p}N_\Lambda M_j(q_j-q_{j}^d) -K^j_{d}N_\Lambda M_j\dot{q}_j$, and
$K^{j}_p \in \mathbb{R}^{n \times n}$ and $K^{j}_d \in \mathbb{R}^{n \times n}$ two symmetric, positive definite matrices. An interesting property of the closed loop system~\eqref{eq:system}--\eqref{eq:torques}--\eqref{posturalNew}--\eqref{eq:forces} is stated in the following Lemma.

\begin{lemma}
\label{lemmaf0}
The closed loop joint space dynamics $\ddot{q}_j$ does not depend upon the wrench redundancy $f_0$.
\end{lemma}

The proof of Lemma \ref{lemmaf0} is in the appendix. 
This result implies also that the linearization of the closed-loop joint dynamics does not depend on $f_0$, thus rendering the gain tuning procedure presented in this paper independent from
the choice of the contact wrenches redundancy. This redundancy is exploited to minimize the joint torques $\tau$ in Eq.~\eqref{eq:torques}.
In the language of \emph{Optimization Theory}, we can rewrite the  control strategy as follows:
\begin{IEEEeqnarray}{RCL}
	\IEEEyesnumber
	\label{inputTorquesSOT}
	f_0^* &=& \argmin_{f_0}  |\tau^*(f_0)|^2 \IEEEyessubnumber  \\
		   &s.t.& \nonumber \\
		   &&Cf_0 < b \IEEEyessubnumber  \label{frictionCones} \\
		   &&\tau^*(f_0) = \argmin_{\tau}  |\tau(f_0) - \tau_0(f_0)|^2 \IEEEyessubnumber	\label{optPost} 
  \\
		   	&& \quad s.t.  \nonumber \\
		   	&& \quad \quad \ \dot{J}(q,\nu)\nu + J(q)\dot{\nu} = 0
		    \IEEEyessubnumber 	\label{constraintsRigid} \\
		   	&& \quad \quad \ \dot{\nu} = M^{-1}(B\tau+J^\top(f_1 + N_{b}f_0) {-} h) \IEEEyessubnumber \\
		   && \quad \quad \ 	\tau_0 = 
		   h_j - J_j^\top(f_1 + N_{b}f_0) + u_0.		    \IEEEyessubnumber
\end{IEEEeqnarray}
The constraints~\eqref{frictionCones} ensure the satisfaction of friction cones, normal contact surface forces, and center-of-pressure constraints. The control torques are then given by $\tau {=} \tau^*(f_0^*)$.

%% file: tex/tuning.tex
\section{GAIN TUNING PROCEDURE}
\label{sec:tuning}

\subsection{Problem Statement}

The goal is to impose desired local properties of the joint dynamics. The choice of focusing on the joint dynamics over other output functions reflects the aim of choosing stiffness and damping at the joint level, without perturbing the task hierarchy of momentum control and stability of the associated zero dynamics via postural control.

Assuming that~\eqref{frictionCones} is always satisfied, the control torques obtained by solving the optimization problem~\eqref{inputTorquesSOT}  depend only on the system state, i.e. 
$\tau = \tau(q,\nu)$. Since it is assumed that the robot stands on (at least) one foot, one can express the system state in terms of  the joint position and velocity, i.e.  $(q,\nu) = (q(q_j),\nu(q_j,\dot{q}_j))$. Then, the joint space dynamics depends only on joint position and velocity, i.e.
$\ddot{q}_j = f(q_j,\dot{q}_j)$ and we can linearize this dynamics about an equilibrium point $(q_j^{d},0)$. 
The process of finding the 
linearized joint dynamics is similar to that presented in~\cite{nava2016}, which yields
\begin{IEEEeqnarray}{RCL}
\ddot{q}_j = -Q_1(q_j -q_j^d) -Q_2\dot{q}_j \IEEEnonumber
\end{IEEEeqnarray}
where $Q_1,Q_2 \in \mathbb{R}^{n\times n}$ are given by 
\begin{IEEEeqnarray}{rCl}
\label{Q1Q2}
Q_1       &=& C_1(q_j^d)K_iC_2(q_j^d)+ C_3(q_j^d)K^{j}_pC_4(q_j^d) \IEEEyessubnumber \label{Q1equat} \\ 
Q_2       &=& C_1(q_j^d)K_pC_2(q_j^d)+ C_3(q_j^d)K^{j}_dC_4(q_j^d)\IEEEyessubnumber
\end{IEEEeqnarray}
and $C_1 =  M_j^{-1}\Lambda^\dagger J_b M_b^{-1}$, $C_2 =  M_b J_b^{\dagger}J_j$, 
$C_3 =  M_j^{-1}N_\Lambda$ and $C_4 = N_\Lambda M_j$. 
Now, let $x \in \mathbb{R}^{2n}$ be defined as follows
$ x := \begin{bmatrix}
        x_1^\top & x_2^\top
       \end{bmatrix}^\top = \begin{bmatrix}
                             q_j^\top - q^{d\top}_j & \dot{q}_j^\top
                            \end{bmatrix}^\top .
$
The linearized joint space dynamics around an equilibrium point ($q_j^{d},0$) is given by
\begin{equation}
    \label{state}
            \dot{x} = \begin{bmatrix}
                \partial_{q_j} \dot{x}_1 & \partial_{\dot{q}_j} \dot{x}_1 \\
                \partial_{q_j} \dot{x}_2 & \partial_{\dot{q}_j} \dot{x}_2
            \end{bmatrix} x = \begin{bmatrix}
                0_n &  1_n \\
               -Q_1 & -Q_2
            \end{bmatrix} x = Ax.
\end{equation}
Then, the optimization problem we attempt at solving is stated next.
\begin{IEEEeqnarray}{RCL}
	\IEEEyesnumber
	\label{problemStatement}
	K_p^{j*},K_d^{j*},K_i^*,K_p^* &{=}& \hspace{-0.45cm} \argmin_{K_p^{j},K_d^{j},K_i,K_p}{\hspace{-0.4cm}|A(K_p^{j},K_d^{j},K_i,K_p){-}A^d|^2} \IEEEeqnarraynumspace \\
		   &s.t.& \nonumber  \\
		   && K_i = K_i^{\top} > 0   \label{constraintsSDP} \IEEEyessubnumber\\
		   && K_p = K_p^{\top} > 0 \IEEEyessubnumber \\
		   && K_p^j = K_p^{j\top} > 0 \IEEEyessubnumber \\ 
		   && K_d^j = K_d^{j\top} > 0 \IEEEyessubnumber
\end{IEEEeqnarray}
where 
$A^d$ is the desired state matrix of the following form:
$A^d =
            \begin{bmatrix} 
                0_n &  1_n \\
               -Q^d_1 & -Q^d_2
            \end{bmatrix} $
with $Q^d_1,Q^d_2 \in \mathbb{R}^{n\times n}$ the desired stiffness and damping matrices. The optimization problem~\eqref{problemStatement}-\eqref{constraintsSDP} may be solved with any nonlinear available optimizer. 
Yet, finding numerical solutions to the optimization problem may be time consuming, which may forbid the on-line use of the optimizer when the desired stiffness and damping are time-varying. In this case, the solutions to~\eqref{problemStatement}-\eqref{constraintsSDP} may also become discontinuous at some time instants.  
We propose below a method for solving on-line the problem~\eqref{problemStatement} that provide continuous solutions for the control gains.

\subsection{Solution to the unconstrained problem}
\label{subsec:uncprob}

Assume that the constraints~\eqref{constraintsSDP} do not hold. When the robot stands on one foot, intuition would suggest that the joint space dynamics can be imposed at will, i.e. there always exist control gains such that the matrices $Q^d_1,Q^d_2$ can be chosen arbitrarily. This section shows, however, that this is not possible because of the two strict stack-of-task  control strategy defined in section~\ref{sec:balancing}.

To show this, we prove that there exist some matrices $Q^d_1$ such that no choice of the control gains renders 
$Q^d_1 = Q_1(K_i,K^{j}_p)$ satisfied.

Now, if  the constraints~\eqref{constraintsSDP} do not hold, then the optimization problem
\eqref{problemStatement} can be rewritten as
\begin{IEEEeqnarray}{RCL}
\label{vectorization}
y_1  &=& \Gamma \begin{bmatrix}
 k_i \\
 k_p^{j}
\end{bmatrix} \IEEEyessubnumber \label{optVecQ1} \\
\Gamma  &=&   \begin{bmatrix}
                          \begin{bmatrix}
                           C_2^{\top} \otimes C_1 
                          \end{bmatrix} &
                          \begin{bmatrix}
                           C_4^{\top} \otimes C_3 
                          \end{bmatrix}               
                   \end{bmatrix} \in \mathbb{R}^{n^2 \times (36+n^2)} \IEEEyessubnumber
\end{IEEEeqnarray}
where $y_1, k_p^j \in \mathbb{R}^{n^2}$, $k_i \in \mathbb{R}^{36}$ are the vectorization of matrices $Q_1, K_p^j$ and $K_i$ obtained by reordering their columns 
into a single column vector, and $\otimes$ the Kronecker product. 
Then, the following result holds.

\begin{lemma}
\label{lemmaGamma}
Assume that $\text{rank}(J) = 6n_c$, and that $n>6n_c$. Then,
the matrix $\Gamma$ is not full row rank, i.e. $\text{rank}(\Gamma) < n^2$.
\end{lemma}

The proof of the above Lemma is in the Appendix. As a consequence of the above Lemma, there exist some desired matrices $Q^d_1$, i.e. $y_1$, such that no control gain implies  the exact solution to $Q^d_1 = Q_1(K_i,K^{j}_p)$. On the other hand, the least square solution to the problem~\eqref{optVecQ1} is given by
\begin{IEEEeqnarray}{RCCCL}
\label{Kronecker}
\begin{bmatrix}
k_i^* \\
k_p^{j*}
\end{bmatrix}   &=& \Gamma^{\dagger}y_1
\end{IEEEeqnarray}

Clearly, the gains $K_i^*,K_p^{j*}$ obtained by the above solution do not in general satisfy the symmetry and positive definiteness constraints~\eqref{constraintsSDP}. A similar procedure can be applied to find the gains $K_d^{j*},K_p^*$ that do not in general satisfy~\eqref{constraintsSDP} but solve in the least square sense $Q^d_2 = Q_2(K_d^{j*},K_p^*)$.
 
\subsection{Enforcing symmetry and positive definiteness constraints}

In the previous section, we solved the problem~\eqref{problemStatement} by assuming that the constraints~\eqref{constraintsSDP} do not hold. Hence, we are now given with a set of gains $K_p^{j*},K_d^{j*},K_i^*,K_p^*$ that may not be symmetric and positive definite. Define $K^{*}~:=~\{ K_p^{j*},K_d^{j*},K_i^*,K_p^*\}$ a matrix of proper dimension. Then,
to enforce symmetry and positive definiteness constraints, we solve (on-line) a second optimization problem for each of the unconstrained optimal gain. More precisely, the problem we solve follows:
\begin{IEEEeqnarray}{RCCL}
\label{optGainsPosDef}
O^*,L^* &=& \argmin_{(O,L)} |K^*-X(O,L)|^2  \\
&\text{s.t.} \nonumber \\
&  OO^{\top} = 1    \IEEEnonumber \\
&  L  &\text{diagonal matrix.} \IEEEnonumber
\end{IEEEeqnarray}
with $X :=O^{\top}\exp(L)O$, $O$  an orthogonal matrix, and $L$  a diagonal
matrix. 
The solution to the problem \eqref{optGainsPosDef} are the matrices $O^{*},L^*$. Therefore, the constrained optimized gain matrix is given by:
\begin{IEEEeqnarray}{RCL}
\label{finalgainOpt}
X^* = O^{*\top}\exp(L^*)O^*
\end{IEEEeqnarray}
Clearly, at this point we just moved the problem from solving the optimization problem~\eqref{problemStatement} with the constraints~\eqref{constraintsSDP} to solving the problems~\eqref{optGainsPosDef} with the constraints of the kind $OO^{\top} = 1$. Now, being $O$ an orthogonal matrix, then $\dot{O} =  OS(v)$, with $v$ a vector of proper dimension depending on the dimension of $O$, and $S(\cdot)$ a skew symmetric matrix. Assuming $v$ and $\dot{L}$ as exogenous control inputs,  one can find Lyapunov-like solutions to the problem~\eqref{optGainsPosDef}. More precisely, the following result holds.

\begin{lemma}
\label{lemmaV}Let $O,L\in \mathbb{R}^{m \times m}$ denote an orthogonal and a diagonal matrix, respectively. Consider the following system:
\begin{IEEEeqnarray}{RCL}
\label{RdotLdot}
\dot{L}              &=&     U      \IEEEyessubnumber \\ 
\dot{O}              &=&   OS(v)  \IEEEyessubnumber       
\end{IEEEeqnarray}
where the vector $v\in \mathbb{R}^{m(m-1)/2}$ and the diagonal matrix $U$ are considered as exogenous control inputs. Define $\tilde{K}~=~K^*-X(O,L)$, and the operator $S^{-1}(v)$ such that $v = S^{-1}(S(v))$. Apply the  control inputs
\begin{IEEEeqnarray}{RCL}
\label{controlInput}
U                 &=&  - K_{U}\exp(-L)\text{diag}(B_1)                \IEEEyessubnumber \\
v                 &=&    K_{v}S^{-1}\left(\tfrac{B_2-B_2^{\top}}{2}\right)         \IEEEyessubnumber  
\end{IEEEeqnarray}
to system \eqref{RdotLdot},
where $K_{U}$ is a positive definite diagonal matrix, $K_{v}$ is a symmetric positive definite matrix, $B_2~=~O^{\top}\exp(L)OK^{*\top}-K^{*\top}O^{\top}\exp(L)O$, $B_1~=~\exp(L)-OK^*O^{\top}$, and $\text{diag}(B_1)$ defined as follows
\begin{IEEEeqnarray}{RCCCL}
\text{diag}(B_1)_{(i,j)} &=& B_{1(i,j)}  &\quad \quad \text{if }& i=j  \nonumber\\
\text{diag}(B_1)_{(i,j)} &=& 0           &\quad \quad \text{if }& i\neq j. \nonumber
\end{IEEEeqnarray}
Then, the following results hold:
\begin{itemize}
 \item If $K^*$ is symmetric and positive definite, the equilibrium point of the closed-loop dynamics $\tilde{K} = 0$ is stable;
 \item The system trajectories $\tilde{K}(t)$ are globally bounded for any $K^* \in \mathbb{R}^{m \times m}$;
 \item $|\tilde{K}(t)| \leq |\tilde{K}(0)|$ for any $K^* \in \mathbb{R}^{m\times m}$.
\end{itemize}
\end{lemma}

\noindent The proof of Lemma \ref{lemmaV} is in the Appendix. 
The above Lemma points out that the \emph{distance} between the optimal, unconstrained solution $K^{*}$ (obtained in Section~\ref{subsec:uncprob}) and the constrained (symmetric, positive definite) gain $X(O(t),L(t))$ is non increasing, i.e. $|\tilde{K}(t)| \leq |\tilde{K}(0)|$. Then, the control laws~\eqref{controlInput} can be viewed as a tracker for symmetric positive definite matrices even when the matrix has to track a non symmetric positive definite matrix (i.e. it does not belong to the same manifold). Let us remark that convergence of the tracking error $\tilde{K}$ to zero is not ensured \emph{a priori}. Simulations we have performed, however, tend to show that the cases when $\tilde{K}$ does not converge to zero are limited, and the analysis on this convergence is currently being developed.

Note also that if the optimal, unconstrained solution $K^{*}$ varies in time slowly, the tracker preserves its properties by continuity. Then, one may think of applying the solution~\eqref{Kronecker}~\eqref{controlInput} on-line for time-varying desired stiffness and damping $Q^d_1(t),Q^d_2(t)$. 
Let us finally observe that we could have avoided to find the intermediate solution~\eqref{Kronecker}, and define the optimization  problem~\eqref{problemStatement} in terms of the parametrization $X(O,L)$, and then apply the procedure explained above to find time evolutions for the constrained gains. Simulations we have performed tend to show that this approach performs worse than the route we propose, and further investigations in this direction are being conducted.

\subsection{Desired matrix correction when two feet balancing}
If the  constraint~\eqref{eqn:constraintsAll} acting on the system represents more than one robot frame fixed with respect to the inertial frame (e.g. two feet balancing), the matrices $Q_1$ and $Q_2$ in~\eqref{Q1Q2} may not be full rank. As a matter of fact, the minimal coordinates describing the constrained mechanical system are fewer than $n$ in this case. Then, the ranks of the desired matrices $Q^d_1$ and $Q^d_2$ must be equal to those of the matrices  $Q_1$ and $Q_2$. In general, the desired matrices $Q^d_1$ and $Q^d_2$  must be corrected according to the constraints acting on the system.

To do this, observe that the \emph{feasible} joint accelerations according to the constraints~\eqref{eqn:constraintsAll} are given by:
\begin{IEEEeqnarray}{RCL}
\ddot{q}_{j} = -\xoverline{J}_j^{\dagger}\dot{\xoverline{J}}_j\dot{q}_{j} + N_{J}\ddot{q}_{j0},
\label{nullAccelerations}
\end{IEEEeqnarray}
where $\xoverline{J}_j = (1_{6n_c}-J_bJ_b^{\dagger})J_j$, $\dot{\xoverline{J}}_j$ is the time derivative of $\xoverline{J}_j$
and $N_{J}$ is the projector onto the null space of $\xoverline{J}_j$. In the above equation, we have used 
$\text{v}_{\mathcal{B}} = -J_b^{\dagger}J_j\dot{q_j}$
and its derivative. Then, given two desired matrices $Q_1^d,Q^d_2$,
we project them as follows
\begin{IEEEeqnarray}{RCL}
\label{correction}
\xoverline{Q}_1^d &=& N_{J}Q_1^d   \\
\xoverline{Q}_2^d &=& N_{J}Q_2^d. \nonumber
\end{IEEEeqnarray}  
to correct the ranks and structure of the desired stiffness and damping according to the constraints~\eqref{eqn:constraintsAll} when $n_c>1$.

%% file: tex/results.tex
\section{SIMULATIONS RESULTS}
\label{sec:results}

\subsection{Simulation Environment}

Simulations are performed on a 23 degree of freedom robot model representing the humanoid iCub. 
The simulation software is developed in MATLAB. The time evolution of the dynamical 
system is obtained through the integration of the system dynamics~\eqref{eq:system} subject to the  constraints  \eqref{eq:constraints_acc}. 
We parametrize the orientation of the base frame using a quaternion representation $\mathcal{Q} \in \mathbb{R}^4 $. The system state is then defined as:
$
\chi :=(
    p_{\mathcal{B}}, \mathcal{Q}, q_j, \dot{p}_{\mathcal{B}}, \omega_{\mathcal{B}}, \dot{q}_j
)
$, and the time derivative is given by
$
\dot{\chi} = (
    \dot{p}_{\mathcal{B}},  \dot{\mathcal{Q}}, \dot{q}_j, \dot{\nu})
$.
The state is integrated through time by means of the numerical integrator MATLAB \emph{ode15s}. 
For the purpose of this paper, both problems~\eqref{Kronecker}~\eqref{controlInput} are solved offline, before starting the state integration. Online implementations of this tuning algorithm will be the subject of a forthcoming publication. 
To integrate the variables $\dot{L},\dot{O}$, we use a  fixed step 
integrator.
The constraints~\eqref{eq:constraints_acc}, as well as $|\mathcal{Q}|=1$ and $OO^{\top}=1$, need to be enforced during the integration phase, and we added correction terms to $\dot{\mathcal{Q}}$, $\dot{O}$.
 \begin{figure}[t]
 \begin{minipage}[c]{8.5cm}
   \centering
   \includegraphics[width=0.8\columnwidth]{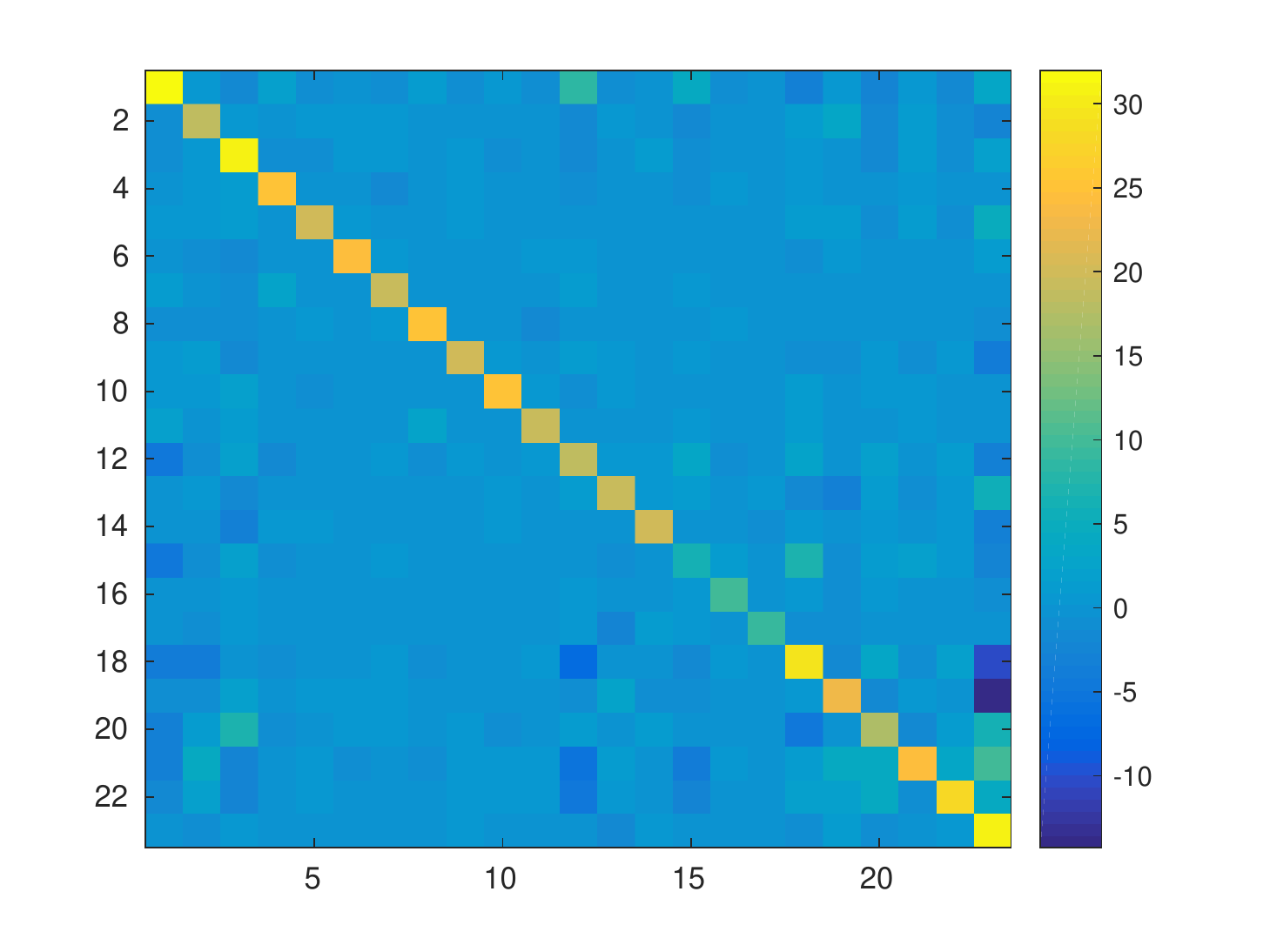}
   \caption{Matrix $Q_1$ after the gain tuning procedure in case of one foot balancing. Simulations run in MATLAB environment.}
    \label{Q1One}
 \end{minipage}
 \ \hspace{2mm} \hspace{3mm} \
 \begin{minipage}[c]{8.5cm}
  \centering
   \includegraphics[width=0.8\columnwidth]{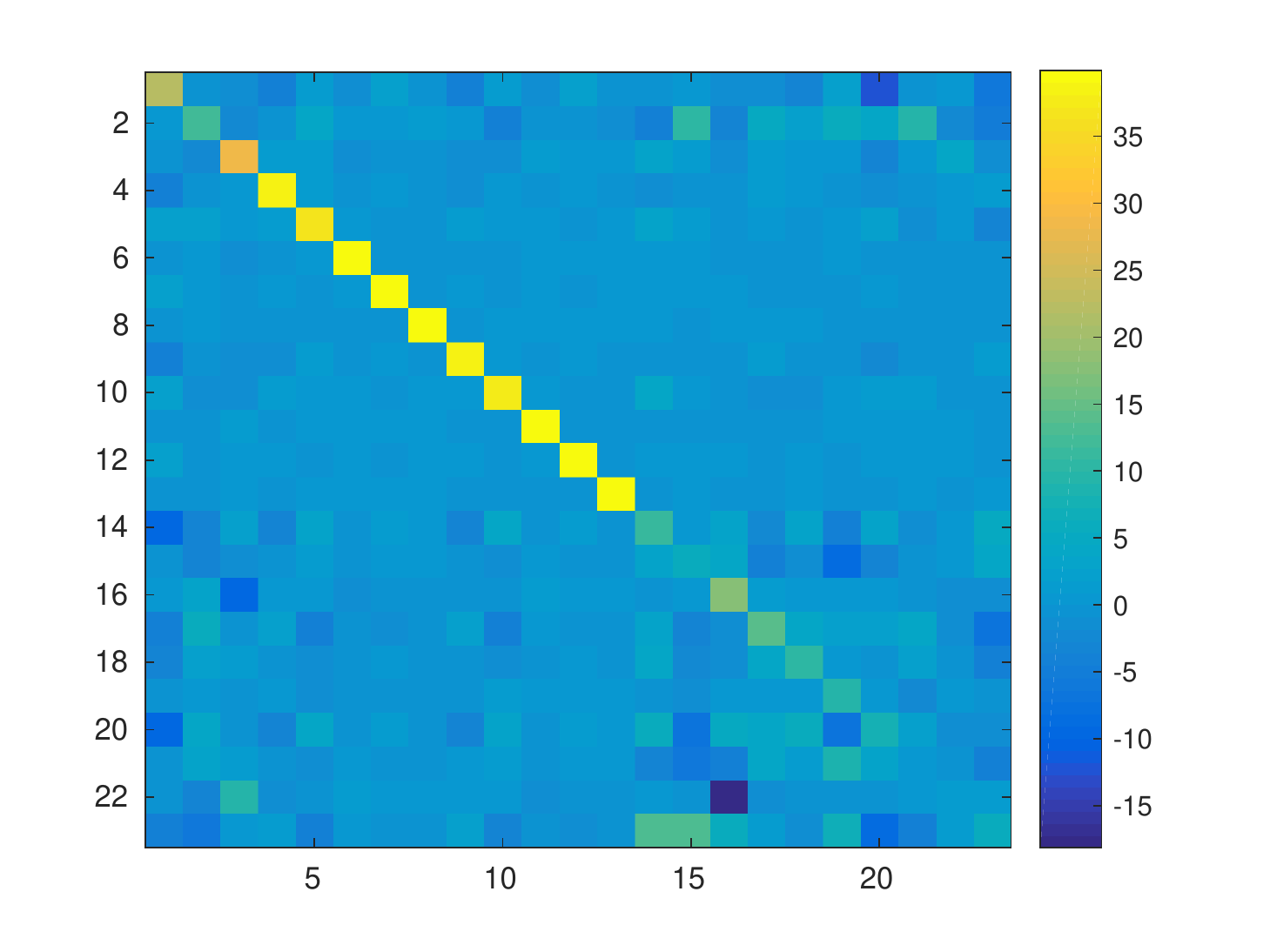}
   \caption{Matrix $Q_1$ after the gain tuning procedure in case of two feet balancing. Simulations run in MATLAB environment.}
    \label{Q1Two}
 \end{minipage}
 \vspace{-0.5cm}
\end{figure}
\subsection{ Results}

Simulations are performed for both the robot balancing on one foot and two feet. We choose
$Q_1^d$ to be positive definite and diagonal, and $Q_2^d = 2\sqrt{Q_1^d}$. 
In this case, the desired joint space dynamics aims at the following properties:
\begin{itemize}
 \item The joint space dynamics be locally decoupled, i.e.
each joint can be tuned separately;
 \item The linearized system is critically damped. This will avoid excessive overshoots in the joint space dynamics.
\end{itemize}
When the robot is balancing on two feet, the matrices $Q_1^d,Q_2^d$  are corrected as in \eqref{correction}.

Figures \ref{Q1One}-\ref{Q1Two} show the shape of matrix $Q_1$ after gain tuning.
Observe that in the case of one foot balancing, the matrix $Q_1$ is close to a diagonal matrix, and this implies that the joint space dynamics is almost locally
decoupled. In case of two feet balancing, it is interesting to observe the effectiveness of the gain tuning procedure by looking at the difference 
between the first 11 rows of $Q_1$, and the last 12 rows, which correspond to the closed chained formed by the legs.

 \begin{figure}[t]
 \begin{minipage}[c]{8.5cm}
   \centering
   \includegraphics[width=.99\columnwidth]{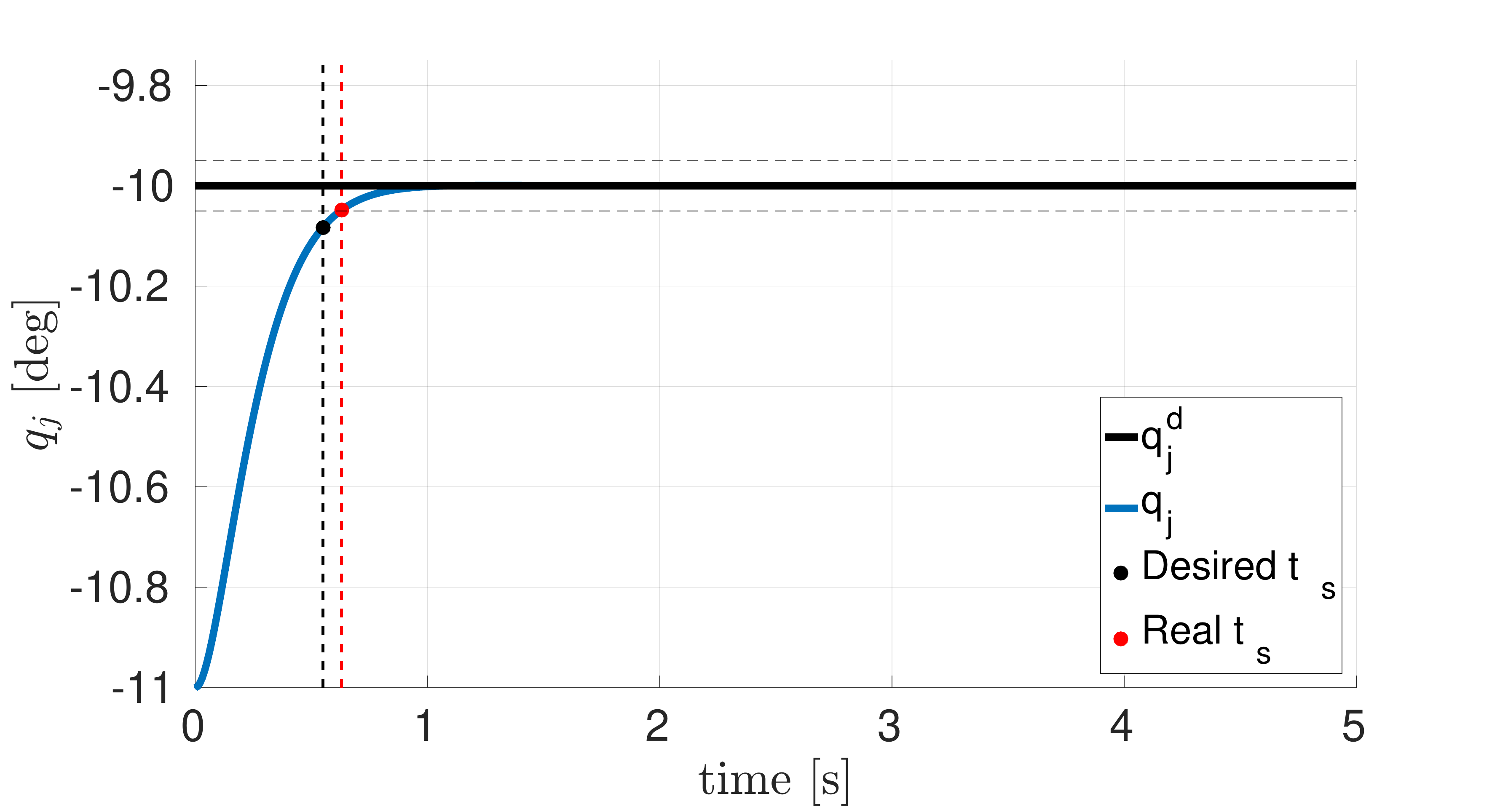}
   \caption{Response of torso pitch to a step input in case of 1 foot balancing. The black dot indicates the desired settling time, while the red dot is the 
            real settling time. The dashed horizontal lines indicate $\pm 5\%$ of the reference position. Simulations run in MATLAB environment.}
    \label{Torso1F}
 \end{minipage}
 \ \hspace{2mm} \hspace{3mm} \
 \begin{minipage}[c]{8.5cm}
  \centering
   \includegraphics[width=.99\columnwidth]{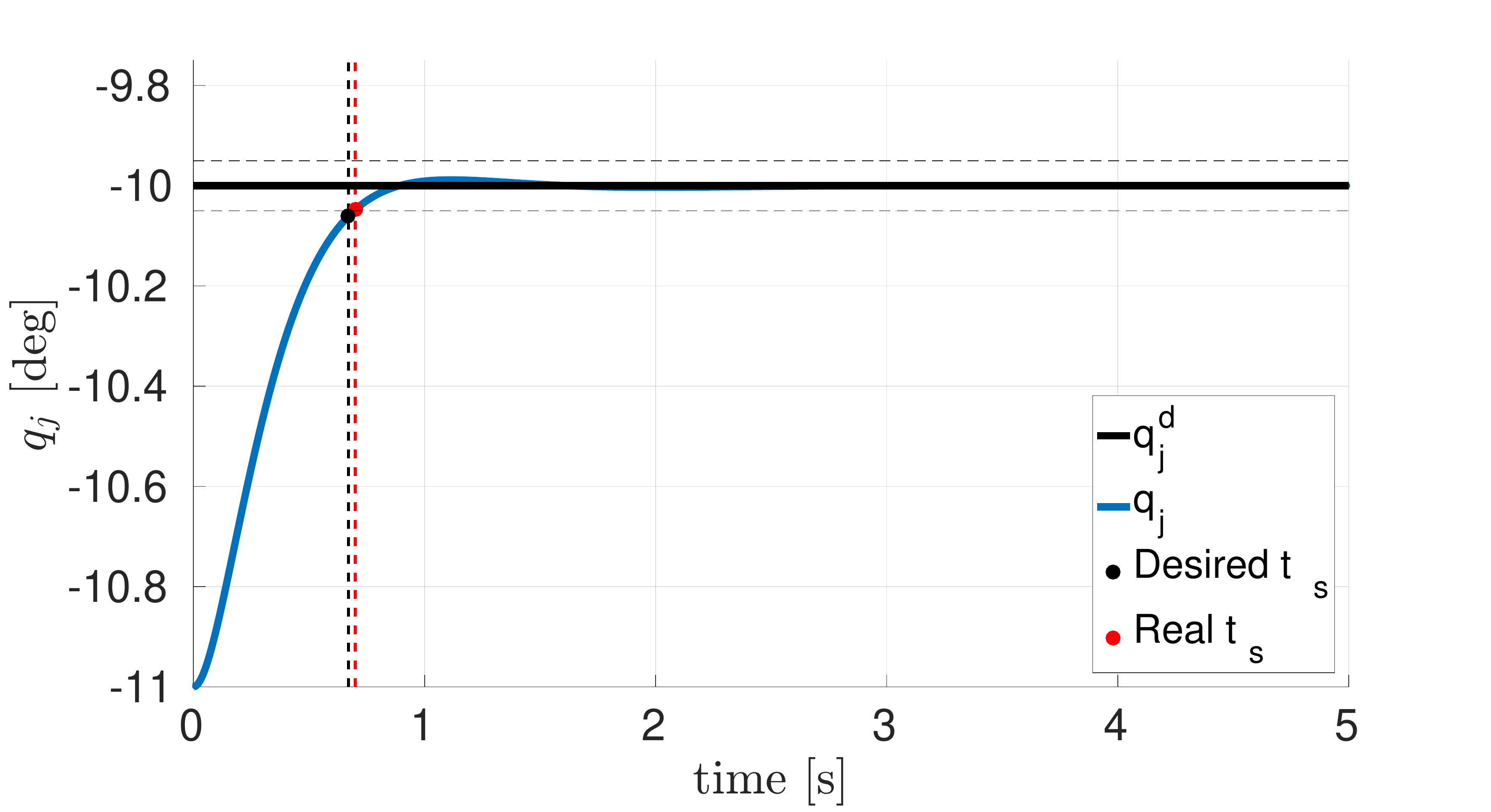}
   \caption{Response of torso pitch to a step input in case of 2 feet balancing. In this case, the real and desired settling time are almost coincident. Simulations run in MATLAB environment.}
    \label{Torso2F}
 \end{minipage}
\end{figure}
To verify that the joint space dynamics is close to the desired dynamics,
we evaluate the dynamical response of each joint to a step input. 
In particular, we focused on the settling time $t_s$. 
It is possible to approximate $t_s$ as 
$t_s \approx \frac{-3}{\real{(\lambda)}}$, 
where $\real(\lambda)$ is the real part of system's eigenvalues. Figures \ref{Torso1F}-\ref{Torso2F} show the dynamics of torso pitch for both one foot and two
feet balancing. In both cases, $t_s \approx t_s^d$, and the system dynamics is then closed to the desired dynamics.

%% file: tex/conclusions.tex
\section{CONCLUSIONS}
\label{sec:conclusions}

This paper has presented a gain tuning procedure for constrained floating base systems controlled through momentum-based control. 
The objective is the achievement of a desired local dynamics for the system's joint space.
The optimization is performed on the linearization around an equilibrium point of the closed-loop system's joint space.  
The constraints on symmetry and positive definiteness of gain matrices are enforced thanks to a tracker for symmetric positive definite matrices. This allows fast resolutions of the constrained optimization problem, which allows one for on-line implementations of the presented algorithms. Simulation results on the humanoid robot iCub show the 
effectiveness of the gain tuning procedure for both the robot balancing on one foot and two feet. 

Further improvements on the gain tuning procedure may be developed in future works.
The gains optimization presented in this paper can be applied in different equilibrium points along a joint reference trajectory, and may be an efficient tuning strategy in case of humanoid walking. On line implementations of the presented algorithm on the real humanoid robot is being investigated.

%% file: tex/appendix.tex
\section*{APPENDIX}
\label{sec:appendix}
\subsection{Proof of Lemma \ref{lemmaf0}}

Recall that
                                        $M$ is block diagonal. 
The joint space dynamics is given by the last $n$ rows of Eq. \eqref{eq:system}:
\begin{equation}
\label{jointSpace}
M_j\ddot{q}_j = J_j^{\top}f - h_j +\tau.
\end{equation}
Moreover, we can rewrite the term $JM^{-1}(h-J^{\top}f)$ in the control torques equations Eq. \eqref{eq:torques} as follows:                
\begin{equation}
    JM^{-1}(h-J^{\top} f) = J_bM_b^{-1}(h_b - J_b^\top f) + \Lambda(h_j - J_j^\top f) \nonumber
\end{equation}
In view of $N_\Lambda = 1_n - \Lambda^{\dagger}\Lambda$ and \eqref{posturalNew}, the control torques~\eqref{eq:torques} become:
\begin{equation}
    \label{newtau}
 \tau = h_j - J_j^{\top}f +\Lambda^\dagger (J_bM_b^{-1}(h_b - J_b^\top f) - \dot{J}\nu) + N_\Lambda u_0
\end{equation}
Substituting Eq. \eqref{newtau} into Eq. \eqref{jointSpace} gives:
\begin{equation}
\label{jointSpaceFinal}
M_j\ddot{q}_j = \Lambda^\dagger (J_bM_b^{-1}(h_b - J_b^\top f) - \dot{J}\nu) + N_\Lambda u_0.
\end{equation}
The only term which contains the wrenches $f$ in Eq. \eqref{jointSpaceFinal} is multiplied by $J_b^{\top}$. Since 
$f = f_1 + N_{b}f_0$, and by definition $J_b^{\top}N_b = 0_{6n_c}$, we have that $J_b^{\top}f  = J_b^{\top}f_1$. Hence, vector $f_0$ does not influence the joint 
space dynamics Eq. \eqref{jointSpaceFinal}.

\subsection{Proof of Lemma \ref{lemmaGamma}}

Given two rectangular matrices $A,B$, recall the properties $\rank{(AB)} \leq \text{min}(\rank{(A)},\rank{(B)})$ and $\rank{(B \otimes A)} = \rank{(A)}\rank{(B)}$, 
where $\rank{(\cdot)}$ denotes the rank of a matrix. We apply the above  properties to evaluate the rank of the matrices $C_1,C_2,C_3,C_4$ in Eq. \eqref{Q1equat}. It
is straightforward to verify that:
$\rank(C_1) \leq 6$, 
$\rank(C_2) \leq 6 $,
$\rank(C_3) \leq n-6n_c$, 
$\rank(C_4) \leq n-6n_c$. It is now possible to evaluate the rank of matrix \[\Gamma = \begin{bmatrix}
                                                                 \begin{bmatrix}
                                                                  C_2^{\top} \otimes C_1 
                                                                 \end{bmatrix} &
                                                                 \begin{bmatrix}
                                                                 C_4^{\top} \otimes C_3 
                                                                 \end{bmatrix}               
                                                             \end{bmatrix} \in \mathbb{R}^{n^2 \times (n^2+36)},\] i.e.
\[\rank(\Gamma) \leq \rank(C_2^{\top} \otimes C_1 ) + \rank( C_4^{\top} \otimes C_3)\leq 36+(n-6n_c)^2. \]
The condition for $\Gamma$ to be full row rank is $\rank(\Gamma) = n^2$. This condition may be verified if $36+(n-6n_c)^2 = n^2$. 
Recall that $n,n_c$ must be positive integers, and that $n > 6n_c$. Assume that $n = 6n_c + k$, with $k \in \mathbb{N}$. Then, one can verify that 
$36+(n-6n_c)^2 < n^2$ yields $36(n^2_c-1) + 12kn_c > 0$, which is always satisfied for any $n_c,k \in \mathbb{N}$. 
As a consequence, $\rank(\Gamma) < n^2$.
\subsection{Proof of Lemma \ref{lemmaV}}

Using Frobenius matrix norms, one has that $ |\tilde{K}|^2 = |K^*-X|^2 = \trace{((K^*-X)^{\top}(K^*-X))}$, where $\trace(\cdot)$ denote the trace operator.
Consider now the candidate Lyapunov function
\begin{IEEEeqnarray}{RCL}
\label{costFunction}
V &=& |\tilde{K}|^2 = \trace{((K^*-X(O,L))^{\top}(K^*-X(O,L)))}
 \IEEEeqnarraynumspace
\end{IEEEeqnarray}
Observe that $V$ is always positive, and $V=0$ iif $\tilde{K}=0$. Then, to prove the three statements in Lemma~\ref{lemmaV}, it suffices to show that $\dot{V}\leq 0$. To do this,
recall that $O$ is an orthogonal matrix, i.e. $OO^{\top} = 1$. Also, observe that
Eq. \eqref{costFunction} can be rewritten as:
\begin{IEEEeqnarray}{RCL}
\label{costFunctionRewr}
V &=& \trace{(K^*K^{*\top}-2K^{*\top}O^{\top}\exp(L)O + \exp(2L))}  \IEEEeqnarraynumspace
\end{IEEEeqnarray}
where we used the properties $\trace(K^{*\top}O^{\top}\exp(L)O) = \trace(O^{\top}\exp(L)OK^*)$
and $\trace(O^{\top}\exp(2L)O) = \trace(OO^{\top}\exp(2L))$.
To compute the time derivative of $V$,
recall that $\dot{O} = OS(v)$, with $S(v)$  a 
skew-symmetric matrix, and $\dot{L} = U$. Then, $\dot{V}$ becomes:
\[\dot{V} = 2\trace(B_1\exp(L)U) + 2\trace(B_2S(v)),\]
with $B_2~=~O^{\top}\exp(L)OK^{*\top}-K^{*\top}O^{\top}\exp(L)O$, and $B_1 = \exp(L)-OK^*O^{\top}$. 
Observe that $\dot{V}~\leq~0$ if both $\trace(B_1\exp(L)U) \leq 0$ and $\trace(B_2S(v)) \leq 0$. Now,
note that $\trace(B_1\exp(L)U)$ can be rewritten as  $\sum_{i=1}^n e_i^{\top}B_1\exp(L)Ue_i$. 
Since the product $\exp(L)U$ is diagonal, then  $\sum_{i=1}^n e_i^{\top}B_1 \exp(L)Ue_i = \sum_{i=1}^n e_i^{\top}B_1e_i\exp(l_i)u_{i}$, where we indicate with $\exp(l_i)u_{i}$ the $i$-th element along the diagonal of $\exp(L)U$.
The trace can then be rewritten as $\sum_{i=1}^n e_i^{\top}B_1e_i\exp(l_i)u_{i} = \sum_{i=1}^n b_{1i}\exp(l_i)u_{i}$ where $b_{1i}$ is the $i$-th element along the diagonal of $B_1$. A possible choice of $u_{i}$ that ensures $\trace(B_1\exp(L)U) \leq 0$ is:
\begin{IEEEeqnarray}{LCL}
\label{dotLi}
u_{i} = -k_{Ui}\exp(-l_i)b_{1i} \quad k_{Ui} > 0.
\end{IEEEeqnarray}
Since $u_{i}$ is the $i$-th element along the diagonal of $U$, \eqref{dotLi} implies $U = -K_{U}\exp(-L)\text{diag}(B_1)$ with $K_{U}$ a diagonal matrix of positive constants.

Now, recall that the matrix $B_2$ can be decomposed as follows:
\begin{IEEEeqnarray}{LCL}
\label{B2parted}
B_2 = \tfrac{(B_2+B_2^{\top})}{2} + S(\omega)
\end{IEEEeqnarray}
where $\omega = S^{-1}\left(\tfrac{(B_2-B_2^{\top})}{2}\right)$. Recall also that the trace of a product between a symmetric and a skew-symmetric matrix is zero. Then, 
$\trace\left(\frac{(B_2+B_2^{\top})}{2}S(v)\right) = 0$. We are now left to evaluate $\trace(S(\omega)S(v))$.
The trace of a matrix product can also be written as $\trace(X^{\top}Y) = \vect(X)^{\top}\vect(Y)$, where $\vect(\cdot)$ is the vectorization operator. Then $\trace(S(\omega)S(v)) = -\trace(S(\omega)^{\top}S(v)) = -\vect((S(\omega))^{\top}\vect(S(v))$. Note that $\vect(S(x)) = Tx \ \forall x$, where the matrix $T$ satisfies  $\tfrac{T^{\top}T}{2} = 1$ due to the skew-symmetry of $S(\cdot)$. Hence, \[\trace(S(\omega)^{\top}S(v)) = -\omega^{\top}T^{\top}Tv = -2\omega^{\top}v\] and this suggests that a possible choice of $v$ is $v~=~K_vS^{-1}(S(\omega)) = K_vS^{-1}\left(\tfrac{(B_2-B_2^{\top})}{2}\right)$.